\documentclass[twocolumn,french,english,prl]{revtex4-1}
\pdfoutput=1
\usepackage[T1]{fontenc}
\usepackage[latin9]{inputenc}
\usepackage{amsmath}
\usepackage{esint}
\usepackage{graphicx}
\usepackage{subfigure}
\usepackage{color}
\usepackage[normalem]{ulem}
\usepackage{amsfonts}

\makeatletter
\@ifundefined{textcolor}{}
{%
 \definecolor{BLACK}{gray}{0}
 \definecolor{WHITE}{gray}{1}
 \definecolor{RED}{rgb}{1,0,0}
 \definecolor{GREEN}{rgb}{0,1,0}
 \definecolor{BLUE}{rgb}{0,0,1}
 \definecolor{CYAN}{cmyk}{1,0,0,0}
 \definecolor{MAGENTA}{cmyk}{0,1,0,0}
 \definecolor{YELLOW}{cmyk}{0,0,1,0}
}

\makeatother

\usepackage{babel}
\makeatletter
\addto\extrasfrench{%
   \providecommand{\fg}{\ifdim\lastskip>\z@\unskip\fi~\frqq}%
}

\makeatother
\begin{document}
\selectlanguage{french}%
\global\long\def\ket#1{|#1\rangle}

\global\long\def\bra#1{\langle#1|}

\global\long\def\id{\mathbb{I}}

\global\long\def\ev#1{\langle#1\rangle}

\global\long\def\ps#1#2{\langle#1|#2\rangle}

\global\long\def\ketbra#1#2{|#1\rangle\langle#2|}

\global\long\def\proj#1{|#1\rangle\langle#1|}

\global\long\def\tr#1#2{\mbox{Tr}_{#2}\left[#1\right]}

\global\long\def\abs#1{\left|#1\right|}

\global\long\def\open#1{\mathring{#1}}

\global\long\def\cE{\mathcal{E}}

\global\long\def\cA{\mathcal{A}}

\global\long\def\cD{\mathcal{D}}

\global\long\def\cP{\mathcal{P}}

\global\long\def\cR{\mathcal{R}}

\global\long\def\cU{\mathcal{U}}

\global\long\def\cH{\mathcal{H}}

\global\long\def\cO{\mathcal{O}}

\global\long\def\cL{\mathcal{L}}

\global\long\def\cB{\mathcal{B}}

\global\long\def\cC{\mathcal{C}}

\global\long\def\cS{\mathcal{S}}

\global\long\def\cN{\mathcal{N}}

\global\long\def\up{\negthinspace\uparrow}

\global\long\def\down{\negthinspace\downarrow}

\global\long\def\pr#1{\mbox{Pr}\left(#1\right)}

\global\long\def\ave#1{\langle#1\rangle}

\global\long\def\expvalue#1#2{\langle#1|#2|#1\rangle}

\global\long\def\matelement#1#2#3{\langle#1|#2|#3\rangle}

\selectlanguage{english}%

\title{Quantitative tomography for continuous variable quantum systems}

\author{Olivier Landon-Cardinal}
\email{olc@physics.mcgill.ca}
\author{Luke C. G. Govia}
\email{govial@physics.mcgill.ca}
\author{Aashish A. Clerk}
\email{clerk@physics.mcgill.ca}

\affiliation{Department of Physics, McGill University, Montr\'eal, Qu\'ebec, Canada.}
\begin{abstract}
We present a continuous variable tomography scheme that reconstructs the Husimi Q-function (Wigner function) by Lagrange interpolation, using measurements of the Q-function (Wigner function) at the \emph{Padua} points, the optimal sampling points for two dimensional reconstruction. Our approach drastically reduces the number of measurements required compared to using equidistant points on a regular grid, although reanalysis of such experiments is possible. The reconstruction algorithm produces a reconstructed function with exponentially decreasing error and quasi-linear runtime in the number of Padua points. Moreover, using the interpolating polynomial of the Q-function, we present a technique to directly estimate the density matrix elements of the continuous variable state, with only linear propagation of input measurement error. Furthermore, we derive a state-independent analytical bound on this error, such that our estimate of the density matrix is accompanied by a measure of its uncertainty.
\end{abstract}
\maketitle

\emph{Introduction}---In modern implementations of quantum information protocols, quantum state tomography~\cite{Hradil97,AJK04} plays a key role. It allows for full characterization of unknown quantum states, as well as verification of prepared resource states. Continuous variable (CV) quantum systems have a wide range of applications in all areas of quantum information, ranging from quantum communication to quantum computing \cite{RevModPhys.77.513}. The need for fast, efficient quantum state tomography of CV systems is heightened by the continuing development of non-classical radiation sources for quantum information \cite{Ourjoumtsev83,Kurochkin:2014aa,Huang:2015aa,Reimer1176,Eichler:2011ab,Zhong:2013aa,Eichler:2014aa,Toyli:aa}, and by recent developments in CV encodings of logical qubits \cite{Mirrahimi:2014aa,Ofek:2016aa,Michael:2016aa}, as well as in quantum simulation with CV systems \cite{Flurin:aa}.

Quantum state tomography of CV systems typically consists of measurement of a quasi-probability distribution, such as the Wigner or Husimi Q-function \cite{RevModPhys.77.513}, from which the density matrix can be reconstructed \cite{RevModPhys.81.299}, and this has been extensively demonstrated in experiment \cite{Kirchmair:2013aa,Wang:2009aa,Vlastakis607,Bertet:2002aa,Deleglise:2008aa,Hofheinz:2009aa,Wang:2016aa}. Unfortunately, full tomography of the Wigner or Q-function is inefficient, as a large number of measurements is required to sample all of phase space. To reduce the number of sampled phase space points, more advanced tomographic schemes involve displacing the state in phase space, and multiple measurements (ideally of the full photon number distribution) at each displacement \cite{Shen:2016aa,Leibfried:1996aa,Hofheinz:2009aa}. For these schemes there is a trade-off between the number of measurement points in phase space, and the number of operator expectation values measured at each point.

In this paper we propose an efficient method to \emph{reconstruct} the full Wigner or Q-function via Lagrange interpolation, using only a small number of measured phase space points. Our method can be applied on any grid of phase space points, thus allowing for reanalysis of previous experiments. However, the ideal points to be measured are known as the Padua points \cite{Caliari2005}, which are the optimal sampling points for two dimensional (2D) interpolation \cite{Bos2006,Caliari2008}. Our scheme requires only a single measurement at each sample point: the Wigner or Q-function value. In addition, we show how individual density matrix elements (including off-diagonal elements) can be estimated from the interpolation reconstruction of the Q-function, such that quantum state tomography of a CV system can be performed directly, without statistical inference.

\emph{Background}---It is well known that the Wigner and Q-function can be measured through a combination of coherent displacements and parity (Wigner) or ideal vacuum (Husimi-Q) measurements \cite{RevModPhys.81.299}, i.e.
\begin{align}
    &W(\alpha) = \frac{1}{\pi}\left[\hat{\Pi}\hat{D}(-\alpha)\rho\hat{D}(\alpha)\right],\\
    &Q(\alpha) = \frac{1}{\pi}\bra{\alpha}\rho\ket{\alpha} = \frac{1}{\pi}{\rm Tr}\left[\ketbra{0}{0}\hat{D}(-\alpha)\rho\hat{D}(\alpha)\right],
\end{align}
where $W(\alpha)$ and $Q(\alpha)$ are the Wigner and Q-function, $\hat{D}(\beta) = \exp\left(\beta\hat{a}^\dagger - \beta^*\hat{a}\right)$ is the usual displacement operator, and $\hat{\Pi} = \left(-1\right)^{\hat{a}^\dagger\hat{a}}$ is the parity operator. Parity measurements can be implemented via interaction with a qubit \cite{Leibfried:1996aa,Wang:2009aa,Vlastakis607,Bertet:2002aa,Deleglise:2008aa,Hofheinz:2009aa,Wang:2016aa}, as can ideal vacuum detection \cite{Kirchmair:2013aa,Flurin:aa}, though this can also be done via photon subtraction measurements \cite{Chen:2011aa,Govia:2012aa,Oi:2013aa}, or heterodyne detection \cite{Kim:1997aa,Eichler:2011aa,Eichler:2012aa}, even of an itinerant state.

\begin{figure*}[ht!]
  \includegraphics[width=0.25 \textwidth]{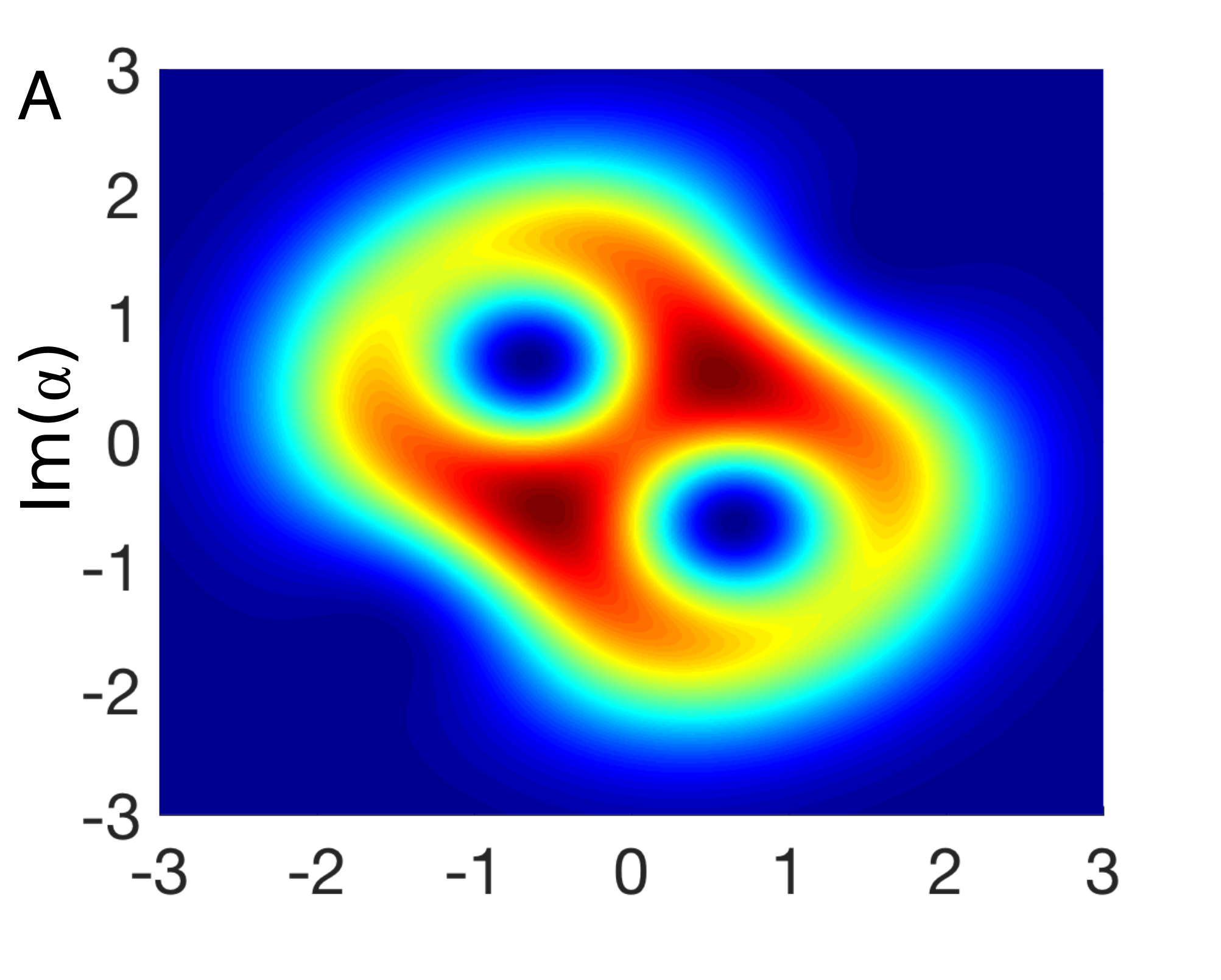}
  \includegraphics[width=0.25 \textwidth]{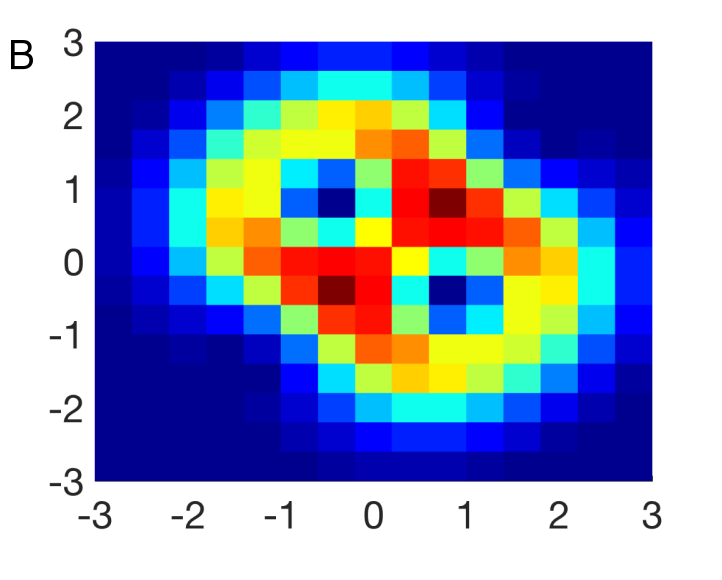}
  \includegraphics[width=0.25 \textwidth]{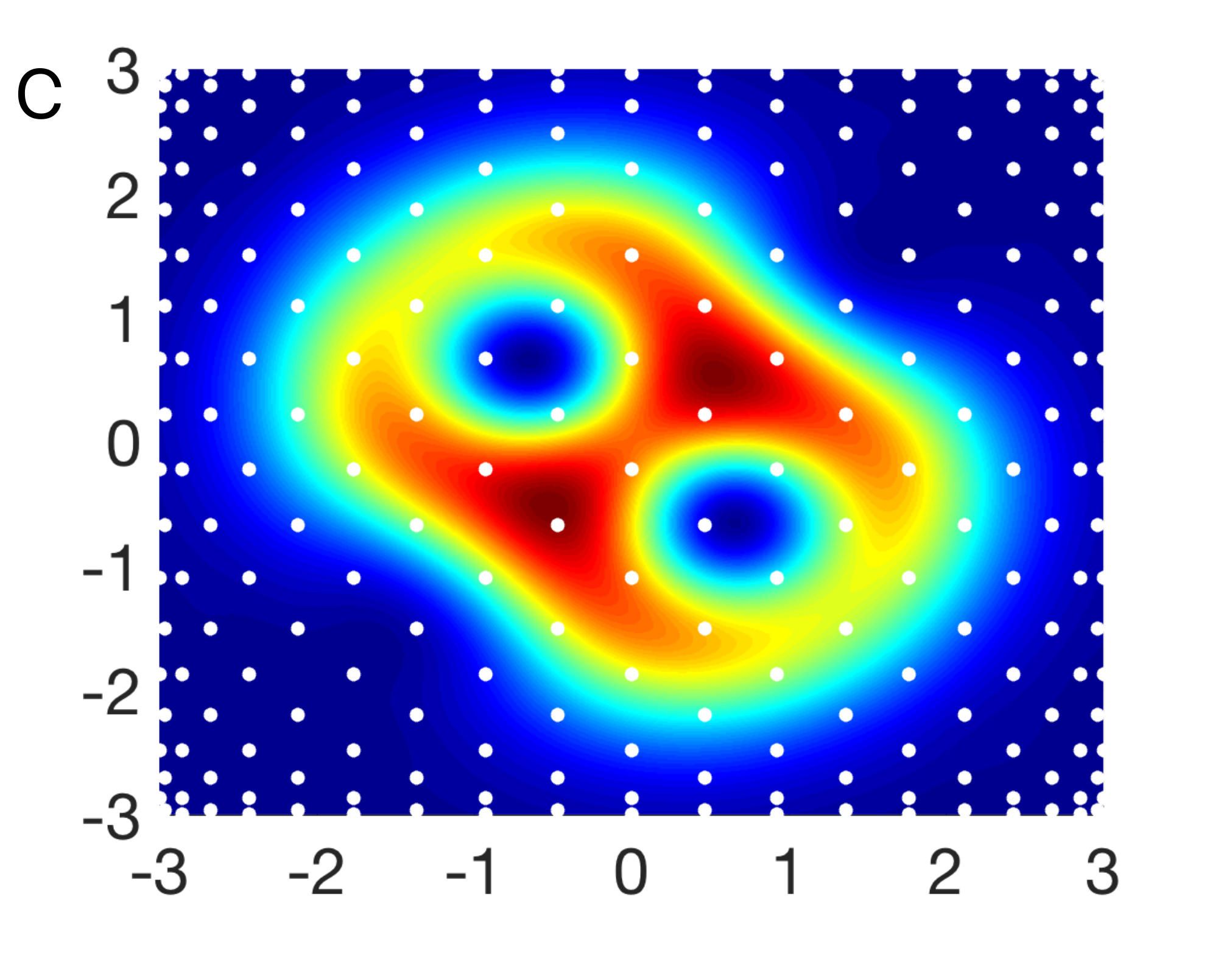}
   \\
  \includegraphics[width=0.25 \textwidth]{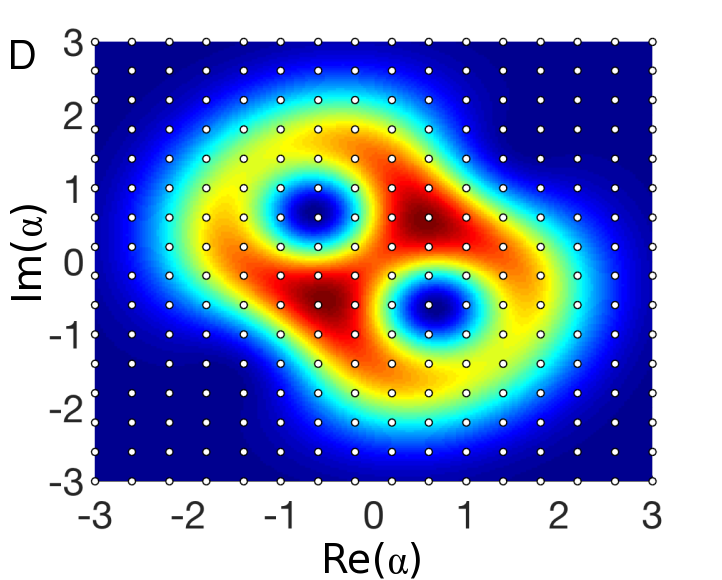}
  \includegraphics[width=0.25 \textwidth]{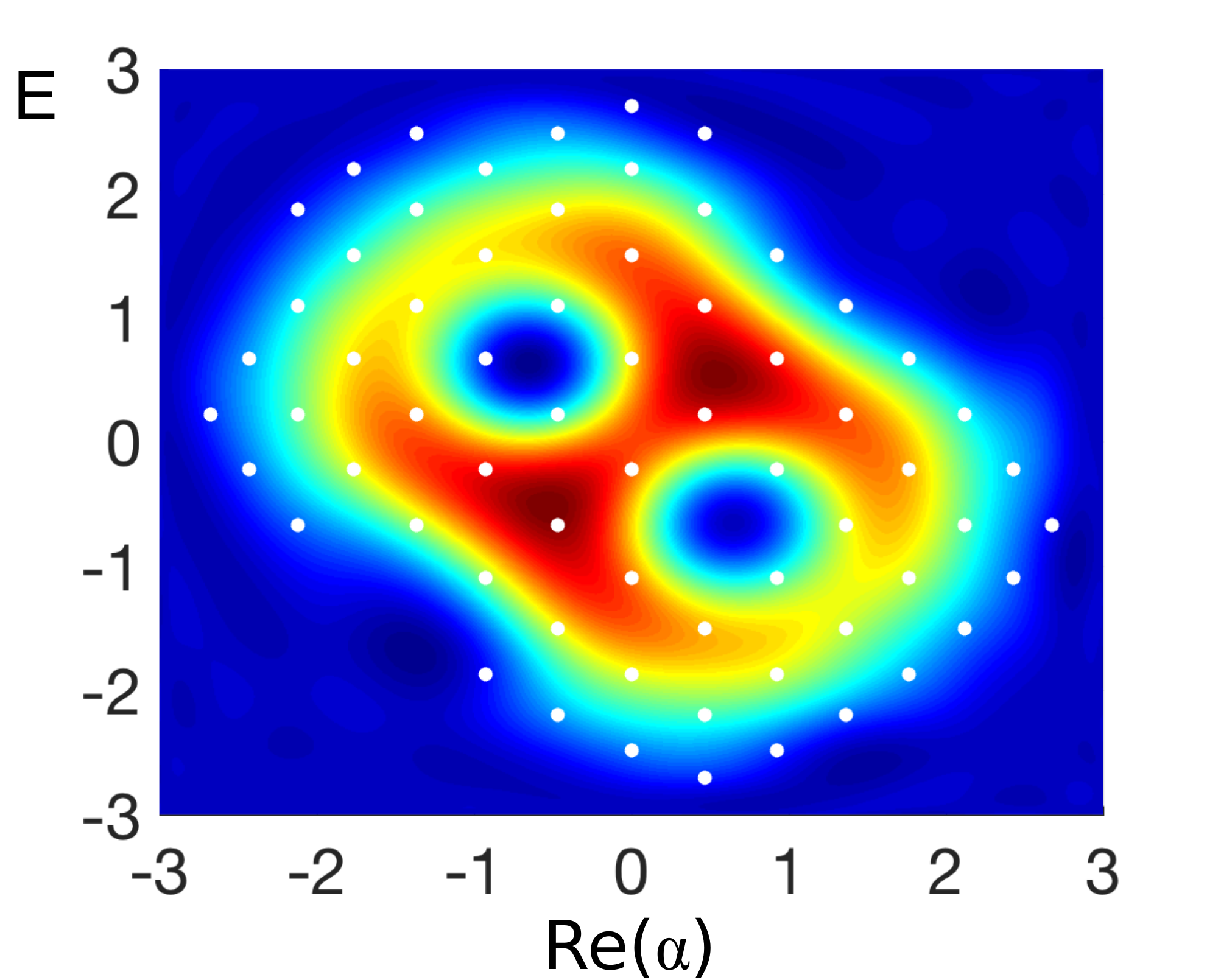}
  \includegraphics[width=0.25 \textwidth]{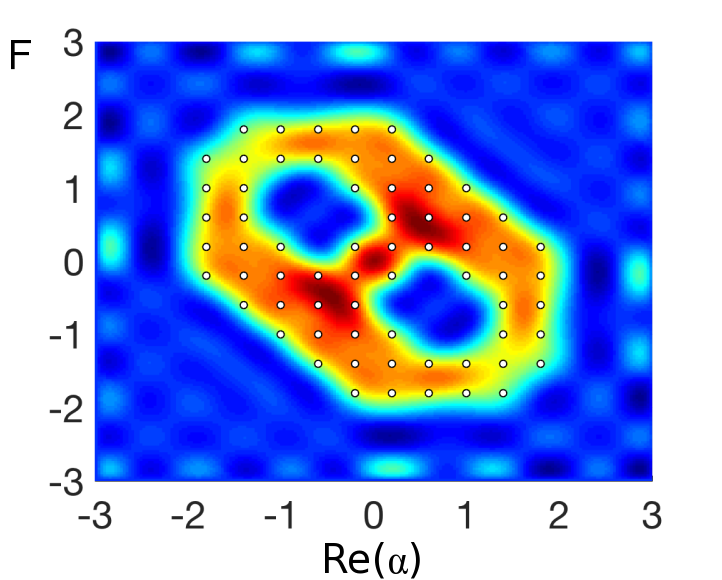}
  \caption{{\bf A} Exact Q-function of  the ideal test state (c.f. Eq.~\eqref{eq:test-state}). {\bf B} Raw data of Q-function measurements taken on a $16\times16$ equidistant grid. {\bf C} A Lagrange interpolation from measurements made at 231 Padua points (white dots). {\bf D} Interpolation on an square grid (white dots) with non-polynomial preprocessing of the data. Both {\bf C} and {\bf D} accurately reconstruct the ideal Q-function of {\bf A}. {\bf E} and {\bf F} show the result of thresholding the interpolations to use only the 65 largest magnitude measurements, in which case the Padua interpolation in {\bf E} outperforms the equidistant grid in {\bf F}.}
  \label{fig:Padua-interpolation}
\end{figure*}

Experimentally, the first step of our method consists of using one of the above techniques to measure the Wigner or Q-function at the Padua points. Using the results of these measurements, one can then reconstruct the full quasi-probability distribution using Lagrange interpolation, and, if the Q-function was measured, directly estimate the density matrix elements of the measured state.

Inspired by the binomial codes of Ref.~\cite{Michael:2016aa}, we use the test state
\begin{align}\label{eq:test-state}
  \ket{\psi}=\tfrac{1}{\sqrt{2}}\left(\ket{\bar{0}}+i\ket{\bar{1}}\right) = \frac{1}{\sqrt{2}}\left(\frac{\ket 0 + \ket 4 }{\sqrt{2}} + i\ket{2}\right),
\end{align}
to illustrate our tomography method, where $\ket{\bar{0}}$ and $\ket{\bar{1}}$ are the logical states for the lowest order binomial code of Ref.~\cite{Michael:2016aa}, with $\ket{n}$ the usual $n$-photon Fock state. This is a reasonable test state as it contain several non-zero density matrix elements of relatively large photon number, and has both real and imaginary off-diagonal elements, such that it tests our reconstruction algorithm under all possible conditions. We will now explain in detail both aspects of our tomography method, quasi-probability reconstruction and density matrix element estimation, and show the results of our method applied on the test state, including the effects of finite measurement error.

\emph{Reconstruction of quasi-probability distributions}---The most na\"ive way to obtain a quasi-probability distribution from measured data is to simply plot the measured values, and increase the number of sampled points until an accurate reconstruction is obtained. In Fig.~\ref{fig:Padua-interpolation}A we show the exact Q-function for our test state, and in Fig.~\ref{fig:Padua-interpolation}B we show the result of plotting 256 sampled points (from an equidistant $16\times16$ grid) of the Q-function. As can be seen, Fig.~\ref{fig:Padua-interpolation}B is a very poor reconstruction of the Q-function, and increasing the number of sampled points to improve the reconstruction requires a considerable increase in experimental resources.

Instead, we propose that the Q-function (or Wigner function) should be reconstructed by Lagrange interpolation. The Q-function in particular is ideal because of its smoothness, as it is known to be an analytic function~\cite{Appleby99}. Generally, interpolation is the problem of reconstructing a function $f$ from the knowledge of its values $\{v_k\}_{k=1}^N$ at $N$ sampling points $\{\alpha_k\}_{k=1}^N$. Using a finite number $N$ of such sampling points allows one to reconstruct the Lagrange polynomial $\mathcal{L}_n [f]$ which obeys
\begin{equation}
 \forall k \; \mathcal{L}_n [f](\alpha_k) = v_k
\end{equation}
where $n$ is the order of the polynomial, with $N={{n+2}\choose{2}}$ for a 2D function.

However, in general, any interpolated function will differ from the polynomial $f^*_n$ (of order $n$) closest to $f$ in uniform norm $||f||_\infty = \sup_x |f(x)|$. The choice of sampling points can strongly affect the quality of the reconstruction~\cite{Berrut2004}, and the quality of the sampling points (more generally of the interpolation scheme $\mathcal{L}_n$) is captured by its operator norm, known as the Lebesgue constant $\Lambda_n$, such that
\begin{equation}
 ||f-\mathcal{L}_n(f)||_\infty \leq (1+\Lambda_n) ||f-f^*_n||_\infty.
\end{equation}
It is well known~\cite{CL09} that in 1D the optimal sampling points are the Chebyshev nodes, whose Lebesgue constant scales logarithmically ($\Lambda_n \sim \log(n)$). On the contrary, equidistant points have a Lebesgue constant exponentially large in $n$, leading to artifact oscillations known as Runge's phenomenon.

The \emph{Padua points} are the 2D equivalent of the Chebyshev points \cite{Caliari2005}, and are currently the best known choice of sampling points for 2D interpolation, since their Lebesgue constant scales as $\log^{2}n$, which is expected to be optimal \cite{Bos2006}. Moreover, the Padua points have a simple generating curve, such that their locations can be efficiently calculated, and allow for efficient computation of the Lagrange polynomial in time $\cO(n^2\log(n))$, using a stable numerical scheme whose open-source Matlab implementation is  available~\footnote{http://profs.sci.univr.it/~caliari/software.htm}. Note that this running time is \emph{linear} in the number of sampling points $N$ (up to logarithmic factors). This Padua interpolation technique is now integrated into standard 2D interpolation packages such as Chebfun2~\cite{TT13}.

In Fig.~\ref{fig:Padua-interpolation}C, we present the reconstructed Q-function for our test state using 231 Padua points (indicated by the white dots), which is an accurate reconstruction of the exact state. It even captures the features of the Q-function close to the origin, despite few sampling points in that region, which we believe is due to the analyticity of the Q-function: points away from the origin constrain the higher-order derivatives of the function, leading to a highly accurate reconstruction.

To compare to the Padua points, in Fig.~\ref{fig:Padua-interpolation}D we consider interpolation on a square grid of 256 points using Chebfun2~\cite{TT13}, which performs non-polynomial pre-processing on the sampled values before producing an interpolating polynomial, in order to reduce the error inherent to equidistant interpolation \footnote{The function chebfun2 from \cite{TT13} for equidistant points first generates a Floater-Hormann interpolant based on rational functions \cite{Floater2007}, and then produces a polynomial approximation of this rational interpolant. This is done to minimize the error inherent to interpolation on an equidistant grid (due to the Runge phenomenon). We compare normal Padua Lagrange interpolation to this hybrid scheme to highlight the power of the Padua points.}. Fig.~\ref{fig:Padua-interpolation}C and Fig.~\ref{fig:Padua-interpolation}D are qualitatively similar, highlighting that any interpolation is a powerful tool. However, there are practical reasons why the Padua points are the better sampling set, as we will now discuss.

Although the theoretical Q-function we aim to reconstruct has only significant values on a rectangle whose main axis is diagonal, we chose to use interpolation points distributed on a square in order to avoid biasing the reconstruction method towards the test state. An alternative scenario is where one is reasonably certain of the state beforehand and require only verification. This can be done by thresholding, where we create a large grid of Padua points, but then only measure at points where we expect (based on our prior knowledge) the quasi-probability to be above some threshold. To illustrate this, we consider a grid of 231 Padua points, and manually set to exactly zero all sampled Q-function values whose absolute value is below $10^{-2}$. The resulting Fig.~\ref{fig:Padua-interpolation}E is very similar to Fig.~\ref{fig:Padua-interpolation}C, but only requires 65 Padua point measurements, which is a significant reduction of experimental effort.

Performing a similar thresholding procedure on the equidistant grid (keeping only the points with the 65 largest values) leads to Fig.~\ref{fig:Padua-interpolation}F, which is clearly a bad reconstruction of the Q-function. The thresholded Padua points cover more of phase space than the thresholded equidistant points, allowing them to more accurately constrain the higher order derivatives of the function, which explains their advantage in verification by thresholding.

The main advantage of the Padua points is that their interpolated function has a quasi-exponentially decreasing error as the number of points is increased, which is not generally true for other point sets (including an equidistant grid).  This greatly benefits the method to estimate the density matrix elements that we now introduce, as it leads to very favorable error scaling, and an analytic bound on the remaining error.

\emph{Direct estimation of density matrix elements}---While quasi-probability distributions give good qualitative descriptions of CV states, for quantitative information one needs the density matrix, which can be calculated using either a linear inversion method, or by statistical inference~\cite{RevModPhys.81.299}. Linear inversion methods are prone to error accumulation, and thus statistical inference methods, such as maximum-likelihood reconstruction, are more commonly employed~\cite{RevModPhys.81.299}. However, with statistical inference it is difficult to assign a measure of confidence or error to the calculated density matrix.

Our method to directly estimate the density matrix elements uses the Lagrange interpolation reconstruction of the Q-function (calculated from measured data), without the need for statistical inference. It does not aim to return a trace one positive semidefinite matrix, but rather estimates the density matrix elements, with calculated bounds on the error in the estimate, based on both input measurement error and the reconstruction error of our algorithm. Our method is related to pattern function reconstruction using the Wigner function \cite{RevModPhys.81.299}, but by using the Q-function, it avoids many of the difficulties inherent to pattern functions.

To calculate the density matrix elements of a state $\rho$ using our method, we first write them as
\begin{align}
  \rho_{jk} = {\rm Tr}\left[\ketbra{k}{j}\rho\right] = \pi\int P_{jk}(\alpha)Q_{\rho}(\alpha)\  d\alpha, \label{eqn:denmatel}
\end{align}
where $Q_{\rho}(\alpha)$ is the Q-function of the measured state, and $P_{jk}(\alpha)$ is the Glauber-Sudarshan P-function representation of the operator $\ketbra{k}{j}$, which in the Fock basis is given by \cite{Sudarshan:1963aa}
\begin{align}
 P_{jk}(\alpha) = \frac{C_{k,j}}{2\pi r} e^{r^2-i(j-k)\theta}\left[\left(-\frac{\partial}{\partial r}\right)^{j+k}\delta(r)\right], \label{eqn:Pfunc}
\end{align}
where $\alpha = re^{i\theta}$ in polar coordinates, and $C_{k,j}=\sqrt{k!\, j!}/\left(2\left(k+j\right)!\right)$, is a combinatorial factor. Eqs.~\eqref{eqn:denmatel}-\eqref{eqn:Pfunc} transform the estimation of the density matrix elements into estimating the derivatives of the sampled Q-function. Inspired from a well-known procedure to optimally estimate finite-differences on an arbitrary grid~\cite{Fornberg81,Fornberg88}, we now show how to relate the Lagrange interpolating polynomial to these derivatives.

The interpolation polynomial can be written in polar coordinates
\begin{equation}
Q_{\rho}(r,\theta)=\sum_{\substack{\phantom{-}0\leq m\leq n\\
-n\leq p\leq n
}
}c_{m,p}r^{m}e^{ip\theta},\label{eq:approx-polar}
\end{equation}
where $c_{m,p}$ are the Q-function coefficients calculated from measured data, and as before, $n$ is the polynomial order of the reconstruction. Using the reconstructed Q-function of Eq.~\eqref{eq:approx-polar}, and the P-function of Eq.~\eqref{eqn:Pfunc} we obtain an expression for the calculated density matrix elements
\begin{equation}
\rho_{jk} = C_{k,j}\sum_{q=0}^{j+k}q!d_{q}^{j+k}c_{q,(j-k)}, \label{eqn:denmatelap}
\end{equation}
where $d_q^{j+k}$ are state-independent constants. Evaluating this expression requires only simple algebra on the coefficients of the reconstructed Q-function, $c_{q,(j-k)}$, which are efficiently calculated from the experimental measurements. We stress that the other factors in Eq.~\eqref{eqn:denmatelap}, including $d_{q}^{j+k}$ and $C_{k,j}$, are state independent constants, and can be efficiently computed once \emph{a priori}.

\begin{figure}[h!]
  \includegraphics[width=0.8\columnwidth]{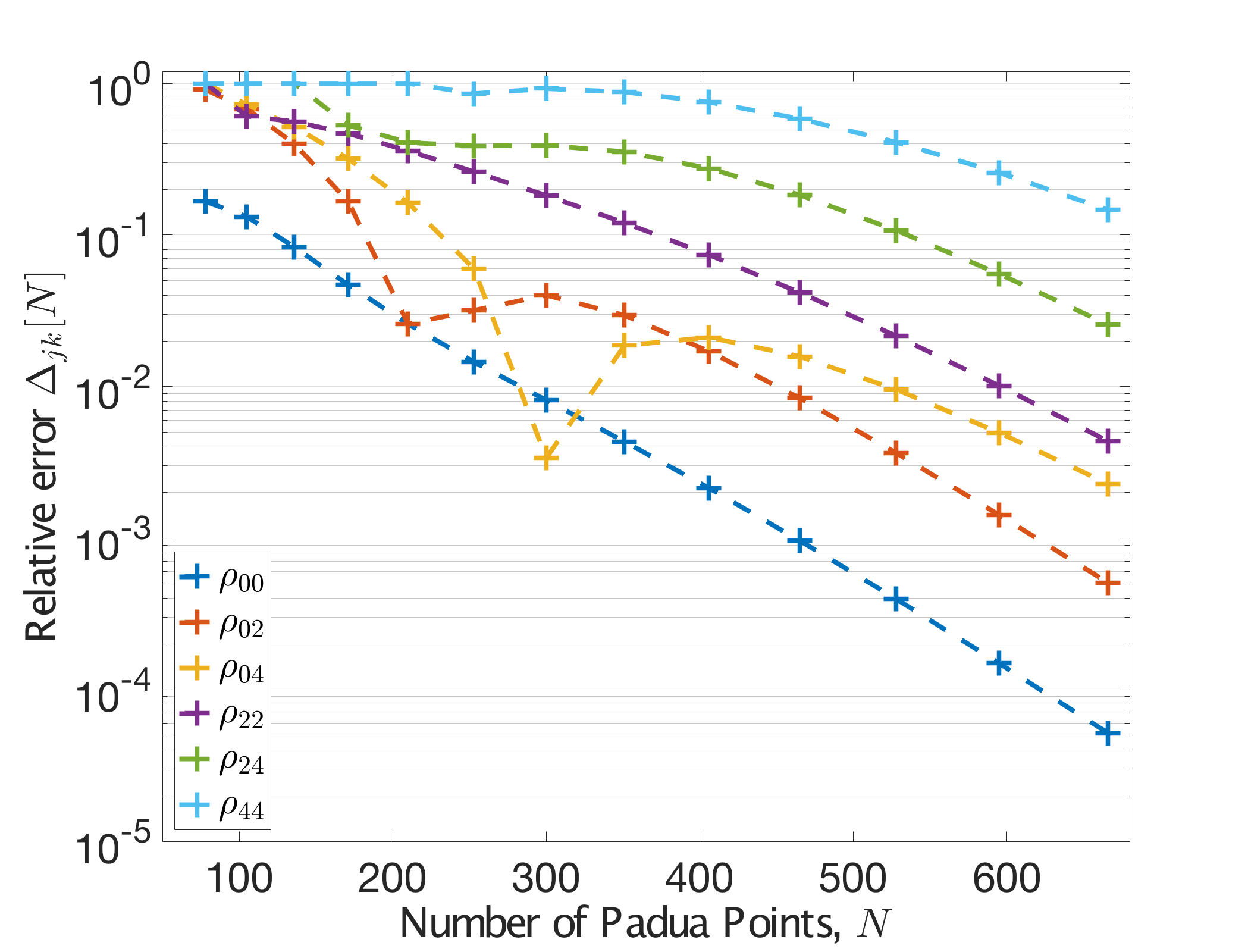}
  \caption{Relative reconstruction error (c.f Eq.~\eqref{eqn:reconE}) as a function of the number of Padua points $N$, for the density matrix elements of the test state of Eq.~\eqref{eq:test-state}. The error decreases exponentially for all density matrix elements as $N$ increases.}
  \label{fig:mean}
\end{figure}

To illustrate our method, we estimate the density matrix of our test state $\ket{\psi}$ from Q-function measurements at the Padua points. To quantify the error intrinsic to Padua reconstruction, we define the \emph{relative reconstruction error}
\begin{align}
  \Delta\rho_{jk}\left[N\right] =  \frac{\abs{\rho_{jk}^{\rm ideal} - \rho_{jk}[N,0]}}{\rho_{jk}^{\rm ideal}}, \label{eqn:reconE}
\end{align}
where $\rho\left[N,0\right]$ is the estimated density matrix element obtained from $N$ Q-function measurements. The relative reconstruction error is plotted in Fig.~\ref{fig:mean} for all density matrix elements of our test state $\ket{\psi}$. As can be seen, the estimated state becomes a better approximation to the actual state as $N$ increases, with the general trend a quasi-exponential decrease in error for large $N$. While our method works for any polynomial approximation of the Q-function, this quasi-exponential scaling is a result of the fact that the Padua points are the optimal reconstruction points in 2D \cite{SupMat}. In Sec. IV of the EPAPS \cite{SupMat}, we show the relative error versus $N$ for an equidistant grid and Chebfun2 interpolation; the error is appreciably worse than for Padua interpolation.

A higher degree polynomial reconstruction (more Padua points) is required to accurately estimate $\rho_{jk}$ for larger $j,k$, as is seen in Fig.~\ref{fig:mean}, where the components with larger $j,k$ have larger relative reconstruction error. Thus, our method is most effective for finite superpositions of Fock states, such as binomial code states \cite{Michael:2016aa}, or states with low average photon number. Alternatively, the center of high quasi-probability can be found using efficient adaptive schemes \cite{Stenberg:2015aa}, and the state displaced to the origin. Our method can then efficiently reconstruct this new displaced state, from which the original state can be calculated.

To accurately reflect imperfect experimental measurements we introduce noise by adding to the sampled Q-function values Gaussian random noise with zero mean and standard deviation $\epsilon$. As this noise is stochastic, each reconstructed element $\rho_{jk}[N,\epsilon]$ will have a mean value $\bar{\rho}_{jk}[N,\epsilon] = \rho_{jk}[N,0]$ and a standard deviation $\sigma_{jk}\left[N,\epsilon\right]$, which are functions of both the number of Padua points $N$, and the noise level $\epsilon$. To examine how input measurement error propagates through our reconstruction algorithm we can use the standard deviation, which we calculate for a range of values of $N$, and input measurement error ranging from $\epsilon=10^{-5}$ to $\epsilon=10^{-1}$ (see \cite{SupMat} for further details). We find linear error scaling
\begin{align}
  \sigma_{jk}\left[N,\epsilon\right] \propto \epsilon,
\end{align}
which could also be surmised from the fact that Eq.~\eqref{eqn:denmatelap} is linear in the Q-function coefficients $c_{q,(j-k)}$ reconstructed by interpolation which we expect to propagate input measurement errors linearly. As such, our reconstruction algorithm does not amplify the input measurement error by more than a constant factor.

Input noise enters Eq.~\eqref{eqn:denmatelap} of our reconstruction algorithm as error in the coefficients $c_{q,(j-k)}$. As all other terms in Eq.~\eqref{eqn:denmatelap} are state-independent, propagation of this error through our algorithm will be state-independent. Therefore, the calculated scaling law of $\sigma_{jk}\left[N,\epsilon\right]$ for our test state is indicative of the scaling law for all states. 

Combining the previous results, our reconstruction scheme gives an estimate $\rho_{jk}[N,\epsilon]$ for the element $\rho_{jk}$ such that
\begin{align}
  \rho_{jk} = \rho_{jk}[N,\epsilon] \pm \Delta_{jk}[N]\rho_{jk}^{\rm ideal} \pm \sigma_{jk}[N,\epsilon] \label{eqn:estimate}
\end{align}
where the first term in Eq.~\eqref{eqn:estimate} accounts for the systematic bias in the estimate due to the absolute reconstruction error, and the second term describes the measurement error due to input noise (assuming one standard deviation error). As this expression shows, our scheme not only calculates the estimated value $\rho_{jk}[N,\epsilon]$, but also estimates the error, given by a linear sum of the absolute reconstruction error and the measurement error.

The input noise, $\epsilon$, can be measured experimentally by benchmarking measurements of known states, and this gives the measurement error, $\sigma_{jk}[N,\epsilon]$, up to a constant factor that can be estimated from Eq.~\eqref{eqn:denmatelap} \cite{SupMat}. The absolute reconstruction error will be state dependent, and can most accurately be calculated by numerical simulation. Alternatively, one can iteratively increase the number of Padua points until $\rho_{jk}[N,\epsilon]$ converges to within $\sigma_{jk}[N,\epsilon]$, which indicates that the measurement error is now dominant and the reconstruction error can be ignored.

\emph{Conclusion}---In this work, we have introduced a continuous variable tomography scheme for accurate and efficient reconstruction of quasi-probability distributions using Lagrange interpolation. The only experimental requirement is the ability to measure the quasi-probability distribution, and if this is done at the Padua points, the optimal sampling points in 2D, our scheme drastically reduces the number of measurements required. Reanalysis of experiments for measurements performed on other point sets is also possible, but forfeits the error bounds guaranteed by using Padua points.

Using the reconstructed Q-function, we have further shown how to estimate the system's density matrix elements, and bounded the error in this estimation. Remarkably, the intrinsic reconstruction error (due to interpolation) decreases exponentially with the number of Padua points, while input measurement error is not amplified by more than a constant factor.

Our scheme will see immediate application in quantum communication and computing protocols with optical or microwave fields, including cavity and circuit QED setups, where it offers significant improvement over the current state of the art. Additionally, our scheme may have application in the characterization of ultra-fast electromagnetic pulses, such as in frequency-resolved optical gating, or other autocorrelation techniques.

\bibliography{ReferencesPPP}

\clearpage

\widetext
\begin{center}
\textbf{\large Supplementary Material: Quantitative tomography for continuous variable quantum systems}
\end{center}

\section{Direct Estimation of Density Matrix Elements}

We start from the expression for the density matrix elements $\rho_{jk}$ used in the main text
\begin{align}
  \rho_{jk} = {\rm Tr}\left[\ketbra{k}{j}\rho\right] = \pi\int P_{jk}(\alpha)Q_{\rho}(\alpha)\  d\alpha, \label{eqn:denmatel_Smat}
\end{align}
where $Q_{\rho}(\alpha)$ is the Q-function of the measured state, and the expression for the Glauber-Sudarshan P-function of $\ketbra{k}{j}$ in the Fock basis \cite{Sudarshan:1963aa}
\begin{align}
 P_{jk}(\alpha) = \frac{C_{k,j}}{2\pi r} e^{r^2-i(j-k)\theta}\left[\left(-\frac{\partial}{\partial r}\right)^{j+k}\delta(r)\right], \label{eqn:Pfunc_Smat}
\end{align}
where $\alpha = re^{i\theta}$ in polar coordinates, and $C_{k,j}=\sqrt{k!\, j!}/\left(2\left(k+j\right)!\right)$, is a combinatorial factor. Substituting the expression for $P_{jk}(\alpha)$ of Eq.~\eqref{eqn:Pfunc_Smat} into Eq.~\eqref{eqn:denmatel_Smat}, we obtain an expression for the density matrix elements in terms of derivatives of the Q-function evaluated at the origin
\begin{align}
\rho_{jk} = C_{k,j}\int_{-\pi}^{\pi}d\theta e^{-i(j-k)\theta}\left[\frac{\partial^{j+k}}{\partial r^{j+k}}\left(e^{r^{2}}Q_{\rho}(r,\theta)\right)\right]_{r=0} = C_{k,j}\int_{-\pi}^{\pi}d\theta e^{-i(j-k)\theta}\sum_{q=0}^{j+k}d_{q}^{j+k}\left[\frac{\partial^{q}Q_{\rho}(r,\theta)}{\partial r^{q}}\right]_{r=0}
\label{eqn:Int_Smat}
\end{align}
where the constant factors $d_{q}^{j+k}$ can be evaluated using the product rule as now follows.

We start by applying the general Leibniz rule
\begin{align}
\frac{\partial^{j+k}}{\partial r^{j+k}}\left(e^{r^{2}}Q_{\rho}(r,\theta)\right)= \sum_{q=0}^{j+k} {{j+k}\choose{q}} \frac{\partial^{j+k-q}}{\partial r^{j+k-q}}\left(e^{r^2}\right)  \frac{\partial^{q}Q_{\rho}}{\partial r^{q}} = e^{-r^2}\sum_{q=0}^{j+k} {{j+k}\choose{q}} (-1)^{j+k-q}H_{j+k-q}(-r)  \frac{\partial^{q}Q_{\rho}}{\partial r^{q}} \label{eqn:Leib_Smat}
\end{align}
where $H_{j+k-q}(-r)$ is the Hermite polynomial of order $j+k-q$. Evaluating Eq.~\eqref{eqn:Leib_Smat} at $r=0$ and comparing to Eq.~\eqref{eqn:Int_Smat}, we obtain
\begin{align}
  d_{q}^{j+k} = (-1)^{j+k-q}{{j+k}\choose{q}}H_{j+k-q}(0)
  =
  \Bigg\{\begin{array}{ll}
  		 (-1)^{\frac{3}{2}\left(j+k-q\right)} \frac{(j+k)!}{q!\left(\frac{j+k-q}{2}\right)!} &  j+k-q {\rm\ even}  \\
  		 0 &  j+k-q {\rm\ odd}
  	\end{array} \label{eqn:d_Smat}
\end{align}
where in the last line we have used the series expansion for the Hermite polynomials, and the fact that odd order Hermite polynomials are zero at $r=0$.

%
%

Finally, using the expression for the reconstructed Q-function in polar coordinates
\begin{equation}
Q_{\rho}(r,\theta)=\sum_{\substack{\phantom{-}0\leq m\leq n\\
-n\leq p\leq n
}
}c_{m,p}r^{m}e^{ip\theta}
\end{equation}
where $n$ is the order of the polynomial reconstruction, we see that the only components that will be survive the integral over $\theta$ in Eq.~\eqref{eqn:Int_Smat} will have $p = j-k$. With this in mind, substituting the reconstructed Q-function into Eq.~\eqref{eqn:Int_Smat} gives the expression for the density matrix elements found in the main text
\begin{equation}
\rho_{jk} = C_{k,j}\sum_{q=0}^{j+k}q!d_{q}^{j+k}c_{q,(j-k)}. \label{eqn:denmatelap_Smat}
\end{equation}

\section{Relative Reconstruction Error Scaling}
\label{sec:RelError}

In analogy to the 1D case \cite{Stewart}, we consider a ``Newton's'' form of the 2D polynomial reconstruction (of order $2n$) of the Q-function, described by the function
\begin{align}
  \mathcal{N}_{2n}(x,y) = \sum^{n-1}_{j=0}a_j\prod_{m=0}^{j-1}(x-x_m)(y-y_m),
\end{align}
where $\left\{(x_m,y_m)\right\}^{n}_{m=0}$ is the set of interpolation points, and the coefficients $a_j$ can be determined iteratively as more points are added to the interpolation set. Adding one additional \emph{arbitrary} point $(x_t,y_t)$ to this set would produce the interpolation polynomial of order $2(n+1)$, and by definition $\mathcal{N}_{2(n+1)}(x_t,y_t) = f(x_t,y_t)$. Therefore, if we wish to calculate the difference between the function and our order $2n$ reconstruction at the arbitrary point $(x_t,y_t)$, we see that it is given by
\begin{align}
f(x_t,y_t) - \mathcal{N}_{2n}(x_t,y_t) = \mathcal{N}_{2(n+1)}(x_t,y_t) - \mathcal{N}_{2n}(x_t,y_t) = a_{n} \prod_{m=0}^{n}(x_t-x_m)(y_t-y_m),
\end{align}
where $a_{n}$ is a function of the whole set $\left\{(x_m,y_m)\right\}^{n}_{m=0}$, and $(x_t,y_t)$. The interpolation error $\abs{\abs{f - \mathcal{N}_{2n}}}_\infty$ is just the maximum of the above expression over all points $(x_t,y_t)$. Choosing the sets $\left\{x_m\right\}^{n}_{m=0}$ and $\left\{y_m\right\}^{n}_{m=0}$ to each be the Chebyshev points, we have that
\begin{align}
  \abs{\abs{\prod_{m=0}^{j-1}(x_t-x_m)(y_t-y_m)}}_\infty = \frac{1}{2^{2n}},
\end{align}
and as such we can bound the interpolation error from above by
\begin{align}
  \abs{\abs{f - \mathcal{N}_{2n}}}_\infty \leq \frac{\abs{\abs{a_n(x_t,y_t)}}_{\infty}}{2^{2n}}.
\end{align}
The best polynomial interpolation of order $2n$, $f^*_{2n}$, will have smaller error than the above expression, and using the Lebesgue constant for the Padua interpolation, we can say that for the Padua interpolation of order $n$ the error will be bounded as
\begin{align}
  \abs{\abs{f - \mathcal{L}^{\rm Pad}_{n}}}_\infty &\leq (1+\log^2(n))\abs{\abs{f - f^*_n}}_\infty \leq (1+\log^2(n))\abs{\abs{f - \mathcal{N}_{n}}}_\infty \leq \frac{(1+\log^2(n))\abs{\abs{a_{\frac{n}{2}}(x_t,y_t)}}_{\infty}}{2^{n}}.
\end{align}
In a 1D reconstruction of Newton's form, the analogue of the function $a_n(x_t,y_t)$ is related to the $(n+1)$ order derivative of the function $f$, which for many functions does not grow exponentially with $n$. If we assume that $a_n(x_t,y_t)$ does not grow exponentially with $n$ in the 2D case, then we obtain a quasi-exponential decrease in the Padua interpolation error.

From Ref.~\cite{Caliari2008}, the error in the reconstructed Q-function coefficients, $c_{m,p}$, is bounded by
\begin{align}
  \delta[c_{m,p}] \leq 4\abs{\abs{f - f^*_n}}_\infty \leq \frac{4\abs{\abs{a_{\frac{n}{2}}(x_t,y_t)}}_{\infty}}{2^{n}}.
\end{align}
Using this and Eqs.~\eqref{eqn:denmatelap_Smat} and \eqref{eqn:d_Smat}, we can derive a bound on the reconstruction error in the calculated density matrix element, $\Delta_{jk}$, in terms of the error in the Q-function coefficients $c_{m,p}$
\begin{align}
  \nonumber\rho^{\rm ideal}_{jk}\Delta_{jk} &= \abs{C_{k,j}\sum_{q=0}^{j+k}q!d_{q}^{j+k}\left(c_{q,(j-k)} - c^{\rm ideal}_{q,(j-k)}\right)} = \abs{\frac{\sqrt{k!j!}}{2}\sum_{\substack{q=0\\{\rm even}}}^{j+k}(-1)^{\frac{3}{2}\left(j+k-q\right)} \frac{\delta[c_{m,p}]}{\left(\frac{j+k-q}{2}\right)!}} \\
  &\leq \frac{\sqrt{k!j!}}{2}\sum_{\substack{q=0\\{\rm even}}}^{j+k} \frac{\delta[c_{m,p}]}{\left(\frac{j+k-q}{2}\right)!} \leq \frac{2\sqrt{k!j!}\abs{\abs{a_{\frac{n}{2}}(x_t,y_t)}}_{\infty}}{2^n}\sum_{\substack{q=0\\{\rm even}}}^{j+k} \frac{1}{\left(\frac{j+k-q}{2}\right)!}
\end{align}
where, as before, $n$ is the order of the Padua reconstruction. By the subscript ``even'' in the summations we mean that the sum is over only values of $q$ such that $j+k-q$ is even. From this expression, we see that if we assume as before that $a_n(x_t,y_t)$ does not grow exponentially with $n$, then we recover the quasi-exponential decay of the reconstruction error seen in the simulated results of Fig.~\ref{fig:mean}.


\section{Scaling of the Standard Deviation of the Measurement Error}

\begin{figure}[h!]
  \includegraphics[width=0.5\columnwidth]{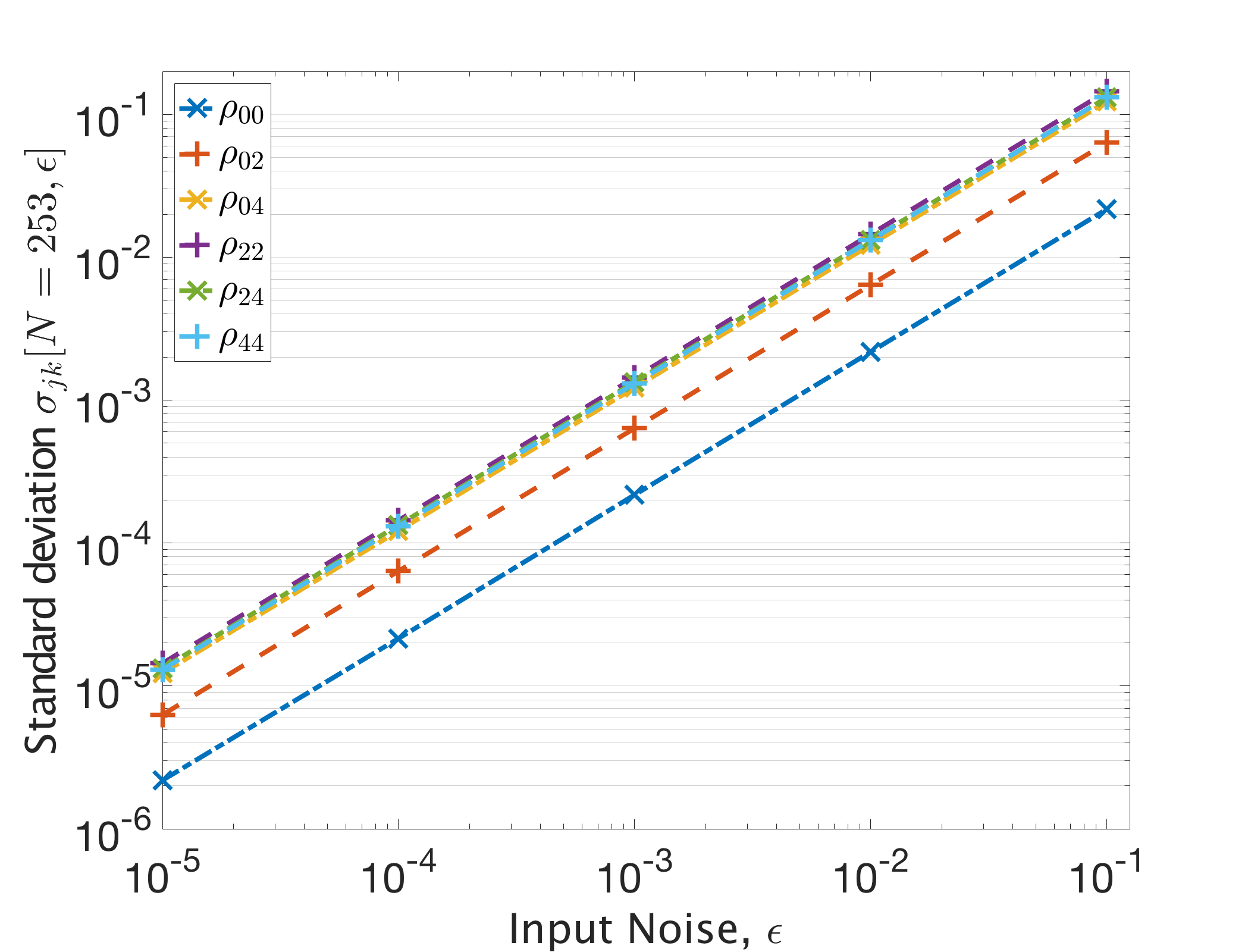}
  \caption{Standard deviation, $\sigma_{jk}[N,\epsilon]$, of the reconstructed density matrix elements as a function of the input noise, $\epsilon$, for a fixed number of Padua points $N = 253$. The slope of each line is one, indicating linear error scaling for all density matrix elements.}
  \label{fig:stddev}
\end{figure}

We calculate the standard deviation $\sigma_{jk}[N,\epsilon]$ for values of $N = (n+1)(n+2)/2$ with $n\in[11,35]$, and $\epsilon \in\{10^{-5},10^{-4},10^{-3},10^{-2},10^{-1}\}$. For all density matrix elements of the test state
\begin{align}
  \ket{\psi}=\frac{1}{\sqrt{2}}\left(\frac{\ket 0 + \ket 4 }{\sqrt{2}} + i\ket{2}\right).
\end{align}
We find that the standard deviation of the reconstructed state scales as
\begin{align}
  \sigma_{jk}\left[N,\epsilon\right] = A_{jk}\epsilon^p,
\end{align}
with $p= 1 \pm 1.6 \times 10^{-3}$, for all $j,k$ and $N$. As an example, we plot $\sigma_{jk}\left[N,\epsilon\right]$ for fixed $N = 253$ as a function of the noise level $\epsilon$ in Fig.~\ref{fig:stddev}, and as can be seen, $\sigma_{jk}\left[N,\epsilon\right]$ increases linearly as $\epsilon$ increases.

Given the form of Eq.~\eqref{eqn:denmatelap_Smat}, a sum of terms linear in $c_{q,(j-k)}$, we would expect linear scaling of $\sigma_{jk}$ with respect to the error in the reconstructed Q-function coefficients $c_{q,(j-k)}$. These coefficients are reconstructed from noisy Q-function measurements, with input measurement error $\epsilon$. Thus, the fact that the standard deviation of $\rho_{jk}$ scales linearly with $\epsilon$ implies that the error in $c_{q,(j-k)}$ must also scale linearly with $\epsilon$. As such, the Q-function reconstruction algorithm does not amplify input measurement error by more than a constant factor.

More concretely, if we assume that the error in the reconstructed Q-function coefficients $c_{q,(j-k)}$ is linear in the error in the Q-function measurements, i.e. it is $K\epsilon$, then we can bound the standard deviation in a similar way as to what was done in section \ref{sec:RelError} for the relative error. Such a procedure results in the expression
\begin{align}
  \sigma_{jk}[N,\epsilon] \leq K\epsilon \frac{\sqrt{k!j!}}{2}\sum_{\substack{q=0\\{\rm even}}}^{j+k} \frac{1}{\left(\frac{j+k-q}{2}\right)!},
\end{align}
which confirms the linear scaling of the standard deviation with $\epsilon$, and bounds the constant factor $A_{jk}$ by
\begin{align}
  A_{jk} \leq K\frac{\sqrt{k!j!}}{2}\sum_{\substack{q=0\\{\rm even}}}^{j+k} \frac{1}{\left(\frac{j+k-q}{2}\right)!}.
\end{align}
It is this bound that allows us to calculate error bars on our density matrix elements in a state independent way, with knowledge of only the measurement error $\epsilon$, and the constant $K$ coming from the Q-function interpolation. In general $K$ is dependent on the number of Padua points, and is difficult to calculate analytically. Heuristically we find that it is a function of both the indices $j,k$ and the number of Padua points $N$, and generally increases as these increase. For the largest $j,k,N$ tested we find $K\simeq40$, though in general $K$ is not this large, and more often $K\simeq1$, such as is the case for the results shown in Fig.~\ref{fig:stddev}.

Note that as we are introducing measurement error artificially, our results may suffer from finite sampling error, as the standard deviation of the reconstructed state is itself a stochastic variable. However, we sample from the noisy distribution sufficient times (on the order of $10^4$) such that the variance of the distribution for the standard deviation of the reconstructed state is several orders of magnitude smaller than the mean value that we report, i.e.  $\sigma_{jk}\left[N,\epsilon\right]$.

\section{Density Matrix Reconstruction with an Equidistant Grid}

\begin{figure}[h!]
  \includegraphics[width=0.5\columnwidth]{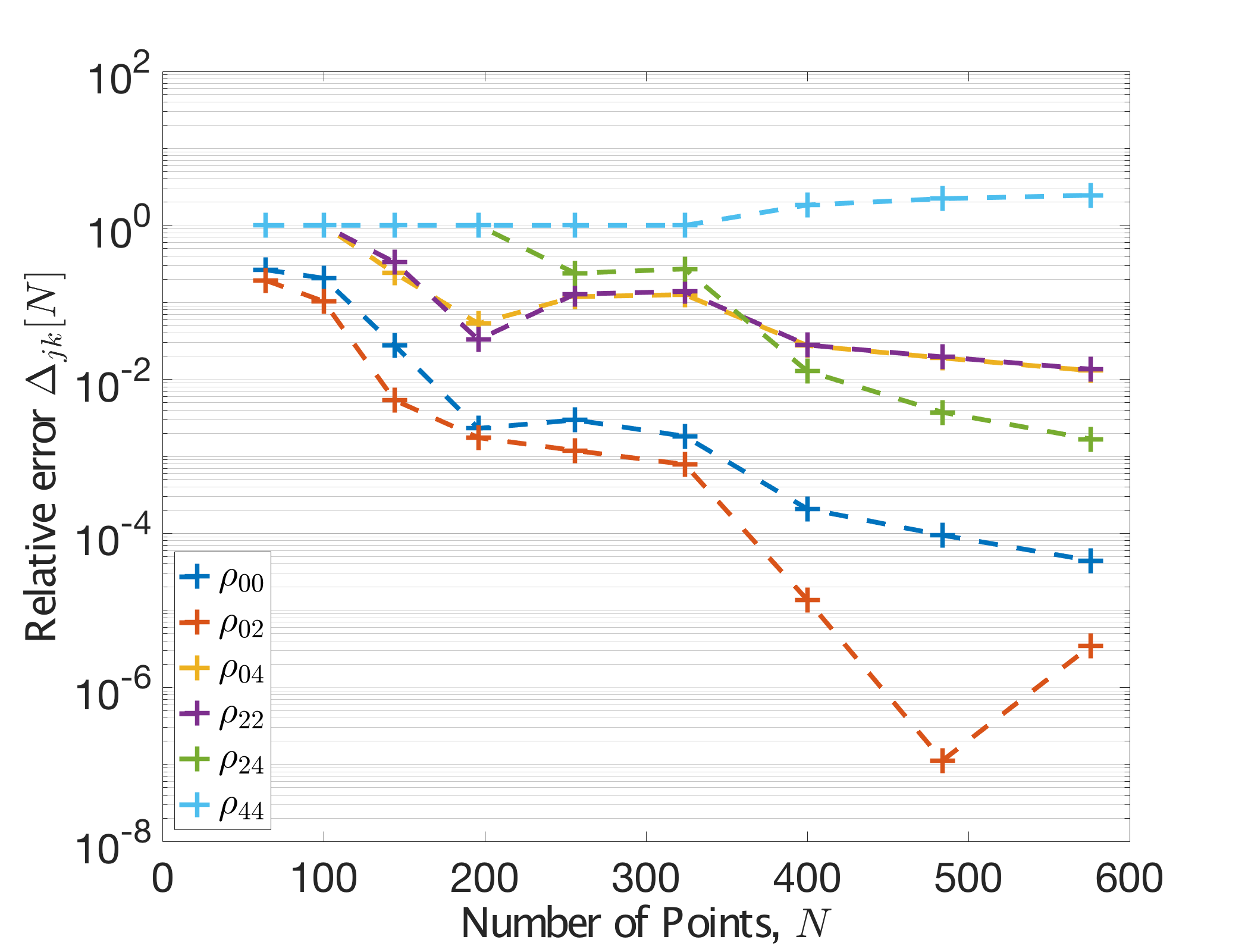}
  \caption{Relative reconstruction error (c.f Eq.~(10) of the main text) as a function of the number of equidistant points $N$, for the density matrix elements of the test state of Eq.~(3) of the main text. The error is not strictly decreasing as $N$ increases.}
  \label{fig:meanE}
\end{figure}

In Fig.~\ref{fig:meanE} we plot the relative reconstruction error for the polynomial approximation generated by Chebfun2~\cite{TT13} on an equidistant grid of points, such as that shown in Fig.~1D of the main text. Despite Chebfun2's non-polynomial preprocessing of the sampled data, which aims to avoid the Runge phenomenon, the reconstruction is considerably worse than when the Padua Lagrange polynomial interpolant is used, as in the main text. In particular, the relative error is no longer strictly decreasing, and appears to saturate for some density matrix elements. This specific example highlights the benefit of performing Lagrange interpolation at the Padua points in comparison to state-of-the-art polynomial approximation on an equidistant grid.

\end{document}